\documentstyle[12pt,aasms4]{article}
\def\kms{\ifmmode {\rm \ km \ s^{-1}}\else $\rm km \ s^{-1}$\fi}
\def\cm{\ifmmode {\rm \ cm }\else $\rm cm$\fi}

\lefthead{Lee and Ahn}
\righthead{Polarization of Ly$\alpha$}

\begin{document}

\title{Polarization of the Ly$\alpha$ line from an anisotropically 
expanding H~I shell in primeval galaxies }
\author{Hee-Won Lee and Sang-Hyeon Ahn}
\affil{Dept. of Astronomy, Seoul National University, Seoul, Korea}
\authoremail{hwlee@astro.snu.ac.kr, sha@astro.snu.ac.kr}

\begin{abstract}
We compute the polarization of the Ly$\alpha$ line photons emerging from an
anisotropically expanding and optically thick medium, which is expected
to operate in many Ly$\alpha$ emitting objects including the primeval
galaxy DLA~2233+131 and Lyman break galaxies.
In the case of a highly optically thick medium, the escape of resonance
line photons is achieved by a large number of resonant local scatterings
followed by a small number of scatterings in the damping wing.
We show that some polarization can develop because
the wing scatterings are coupled with strong spatial diffusion
which depends on the scattering geometry and kinematics.
The case of a slab with a finite scattering optical depth and
expansion velocity of $\sim 100~\kms$ is investigated
and it is found that Ly$\alpha$ photons are emergent with the linear degree
of polarization up to 10 per cent when the typical scattering optical depth
$\tau {\gtrsim} 10^5$. We subsequently investigate the polarization
of Ly$\alpha$ photons emerging from a spherical shell 
obscured partially by an opaque component and we obtain $\sim$ 5 per cent
of polarization. It is proposed that a positive detection of polarized
Ly$\alpha$ with P-Cygni type profile from cosmological objects can be a
strong test of the expanding shell structure obscured by a disk-like
component.
\end{abstract}


\section{Introduction}

Various astronomical objects in the cosmological scales show
P-Cygni type profiles in the Ly$\alpha$ emission.
These objects include the most remote galaxy at $z=4.92$
gravitationally lensed by CL1358+62 \markcite{fra97} (Franx et al. 1997),
high $z$ galaxies observed with the {\it Hubble Space Telescope} and the
Keck telescopes \markcite{ste96a, ste96b, gia96, low97} 
(Steidel et al. 1996, Giavalisco et al. 1996, Lowenthal et al. 1997) 
and the damped Lyman $\alpha$ (hereafter DLA) candidates
\markcite{djor96, djor97} (Djorgovski et al. 1996, 1997).
Similar P-Cygni Ly$\alpha$ profiles are found in nearby starburst galaxies,
which are sometimes classified as Wolf-Rayet galaxies, blue compact galaxies,
or H~II galaxies \markcite{kun96, hec97, sah97, leq95, leg97}
 (e.g. Kunth et al. 1996, Heckman and Leitherer 1997,
Sahu and Blades 1997, Lequeux et al. 1995, Legrand et al. 1997, etc.).
The column density $N_{HI}$ of neutral hydrogen in these systems is
usually found to be in the range $N_{HI}\sim 10^{19-21}~\cm^{-2}$.
 
The primeval galaxies or the first star clusters expected to be found
at $z>5$ epoch may possess a central super star cluster
surrounded by neutral hydrogen of high column density
\markcite{hai97a, hai97b} (Haiman and Loeb 1997a,b).
These surrounding layers can be accelerated by the
expanding H~II region just outside the super star cluster.
It is hoped that in the near future with the advent of the {{\it Next
Generation Space Telescope} (NGST), the infrared spectra of
these infant galaxies will be accessible and that the
observational confirmation of the ubiquity of P-Cygni type Ly$\alpha$ 
profiles may test the above hypothesis.

\markcite{AL98} Ahn and Lee (1998, hereafter AL98) investigated the
Ly$\alpha$ line formation in a thick and expanding medium.
It was emphasized that the profile formation should be studied by
accurately computing the contributions from photons back scattered 
by receding medium and wing-scattered photons \markcite{leg97}
(see also Legrand et al. 1997).
It is well known that the properties of the Ly$\alpha$ photons
scattered in the damping wing are characterized by the Rayleigh phase
function \markcite{ste80} (e.g. Stenflo 1980). 
This is in contrast with the degree of polarization
$p=0$ resulting from a resonance transition between $1S_{1/2}$ and $2P_{1/2}$
and $p=3/7$ obtained for the $1S_{1/2}$ and $2P_{3/2}$ transition
\markcite{lee94b} (Lee et al. 1994).
 
In a moderately thick and static medium a negligibly polarized
flux is expected because the photons are locally scattered many times
and get isotropized before they escape to the observer
\markcite{lee94b} (e.g. Lee 1994). 
However, in a very thick medium, the escape is achieved
by a large number of local resonant scatterings followed by
a small number of scatterings in the damping wing.
Hence, in the wing regime
the spatial diffusion becomes important and the radiation field
may get anisotropic depending on the scattering geometry.
Therefore, the emergent line photons may get polarized and
also anisotropic kinematics introduced in the medium can
enhance the polarization.
 
In this {\it Letter}, we compute the polarized flux
of the emergent Ly$\alpha$ from an optically thick and expanding slab.
This result is applied to a hemi-spherical shell that is expected
in various systems including primeval galaxies
exhibiting P-Cygni profiles.
 
\section{Theory}
\subsection{Anisotropically Expanding Slab}

In AL98 is given a brief discussion about the line radiative
transfer in an expanding and optically thick medium 
\markcite{ost62, ada72, ryb78, sob60}
(see also Osterbrock 1962, Adams 1972, Rybicki and Hummer 1978, Sobolev 1960).
We describe the model adopted in this work and introduce the computation
procedures.
 
First, we consider a slab expanding in the normal direction, for which
the bulk velocity law is given by ${\bf V}(z) = H z \hat{\bf z}$
where $z-$ direction is chosen to coincide with the normal direction and the
velocity gradient $H$ is constant. Here, the coordinate system is chosen
such that at $z=0$ the bulk velocity vanishes.
We also assume that the medium is characterized by a uniform density and a
fixed temperature.
 
The scattering optical depth of a given line photon with frequency
deviation $x$ in units of the Doppler width is given by
$$
\tau(s) =
{\tau_0 \over \cos^2\theta}
 \int_{-\infty}^{\infty} du \ e^{-u^2}
\left[ \tan^{-1}\left( {{x-u} \over a} \right) - \tan^{-1}\left(
{x-u-s\cos^2\theta \over a} \right)\right] ,  \eqno(1)
$$
where the parameter $s$ is used for the path length $l$ of the
photon by the relation $s\equiv H l/v_{th}$, and
$\theta$ is the angle between the $z-$ axis and the propagation
direction of the photon.  Here,
$\tau_0={{\pi e^2}\over {mc}}f_{osc}{n_0\over \pi^{3/2}}{\lambda_0\over H}$
is the Sobolev type scattering optical depth.
We fix the velocity gradient $H$, and the bulk velocity difference
at the bottom and the top of the slab is given by
$$\Delta V= H d = 100\left({s_{max}\over 10}\right)~\kms, \eqno(2)$$
where $d=s_{max} v_{th}/H$ is the slab thickness.
Other parameters have their usual meanings 
\markcite{AL98} (see AL98 for detail).
 
The polarization state of a given photon is described by a
density matrix and the circular polarization is zero
in this work because of the azimuthal symmetry.
The density matrix associated with the scattered Ly$\alpha$ photon is
dependent on the scattering type and in particular
if the scattering occurs in the damping wing, then the phase matrix
is identical to that for the Thomson scattering or the classical
Rayleigh scattering. A more detailed description of the density
matrix associated with the
scattered photon as a function of the incident photon states
is provided by \markcite{ahn98} Ahn \& Lee (1998).
 
We locate the incident photon source at the bottom of the slab and
perform a Monte Carlo simulation to get the emergent flux from the
other side of the slab. The initial incident profile
is chosen to be a Gaussian of width $\sigma=5 v_{th}$.
The linear degree of polarization is computed
with respect to the normal direction of the plane spanned by the wave
vector $\hat{\bf k}$ of the emergent photon and $z$-direction. 
In this work a positive linear degree of polarization 
represents the electric vector lying in that direction and a negative polarization 
perpendicular to it.

\subsection{Expanding Hemispherical Shell}

The preceding slab analysis is applied to deal with an expanding spherical shell.
Assuming that the spherical shell can be decomposed into slabs
with normal vectors distributed spherically,
we integrate the emergent flux from each slab component to get the profiles 
and the polarization. 
The hemispherical geometry is chosen because it 
describes shell structures in many objects including neutral shells 
surrounding an HII region such as NGC 1705 \markcite{meu95} (Meurer et al. 1995),
or galactic supershells 
\markcite{kam91, hei79, hei84, koo91, koo92, rea93}
(e.g. Kamphuis et al. 1991, Heiles 1979, 1984, Koo 1991,
1992, Reach 1993) possibly obscured by optically thick dust lanes
\markcite{ich95, sco98} (Ichikawa et al. 1994, Scoville et al. 1998). 

\section{Results}
\subsection{Ly$\alpha$ Line Formation and Polarization}

\placefigure{fig1}

In Fig.~1 we show the main results. We collect those photons emerging with
$\mu \equiv \cos \theta = 0.5\pm 0.05$.
The overall polarization increases as $\Delta V$ increases. The polarized flux is
significant in the red part when $\tau_0=10^4, s_{max}=10$ and
$\tau_0=10^5, s_{max}=1$. A typical degree of polarization
is found to be 10 per cent in those cases, and the polarization
direction is perpendicular to the normal direction of the slab. 
We note that only a small degree of polarization is obtained when
$\tau_0\le 10^3$, where scatterings rarely occur in the damping
wing regime.

Line photons with moderate frequency deviation ($|x| {\lesssim} 4$)
suffer a large number of resonance scatterings at a local
site and the radiation field gets isotropized 
\markcite{lee94b} (e.g. Lee 1994).
After a severe frequency diffusion, scatterings occur dominantly in 
the wing just before escape.
Therefore, diffusion in the real space as well as in frequency space
is now important, and thereby the escape of the line photons
is dependent highly on the detailed geometry
of the scattering medium. Thus, being 
characterized by the Rayleigh phase function, the wing scattered photons 
can be significantly polarized. 
Furthermore, an introduction of an anisotropic expansion enhances
spatial excursions into the normal direction and leads to
an increased degree of polarization.

In a thin electron scattering atmosphere, the
emergent photons are polarized in the normal direction to
the slab plane \markcite{phi86} 
(see Phillips \& M\'eszar\'os 1986), and we point out
that a similar analogy explains the polarization direction
of the emergent Ly$\alpha$ photons shown in Fig.~1,
because they suffer a few damping wing scatterings
just before escape. 

In the case of semi-infinite electron scattering atmosphere
of which a detailed investigation was made by \markcite{cha60} Chandrasekhar (1960),
the emergent continuum radiation is polarized in
the slab plane direction, because the electric vector associated
with the emergent radiation relaxes to the slab plane direction.
A similar phenomenon is expected in the case where photons
suffering a large number of damping wing scatterings just before
escape are dominantly obtained. However, due to frequency redistribution 
they are distributed in a large frequency range and mixed with other 
line photons suffering fewer wing scatterings. Hence only near the line center
a small flux of photons with a large number of wing scatterings
can be found. This observation leads into a possible polarization
flip near the line center, which is marginally seen in Fig.~1
in the cases $\tau_0=10^4, s_{max}=10$ and $\tau_0=10^5, s_{max}=1$. 

A more detailed analysis about the basic physical processes will be
discussed in a subsequent paper \markcite{ahn98} (Ahn and Lee 1998).

\subsection{Hemispherical Shell}

\placefigure{fig2}

In Fig.~2 we show the averaged degree of polarization
expected from a hemispherical shell described in section 2.2.
Here, the obscuration by an optically thick dust lane is assumed to 
extend from the equatorial region to the region with colatitude $\theta=
0^\circ, 30^\circ$ and $60^\circ$.
The adopted paramteres are $\tau_0=10^4, s_{max}=10$.

Significant polarization and polarized flux are found in the red part
and this depends on the obscuration fraction.
In the case of full obscuration ($\theta = 0^\circ$)
we obtain negligibly small polarization.
The linear degree of polarization is obtained $\sim$ 5 per cent where
the obscuration fraction is 1/2.  Therefore, a positive detection of polarized
Ly$\alpha$ with P-Cygni type profile from a cosmological object can be a 
strong test of the expanding shell structure with obscuration.

\section{Summary and Discussion}

It seems a general consensus that the P-Cygni type Ly$\alpha$ emissions
are originated from expanding envelopes of H~II regions,
which are indicative of the massive star formation.
These are often obscured by dust lanes or thick molecular
disks \markcite{ich94, sco98} (Ichikawa et al. 1994, Scoville et al. 1998). 

In this {\it Letter} we computed the polarization of the Ly$\alpha$ photons
that are transferred through an optically thick and expanding neutral 
hydrogen layer. Anisotropic expansion and high column density are coupled to
enhance scatterings in the damping wing into the direction corresponding to
the largest velocity gradient, which results in highly polarized
emergent flux. In particular, in a spherical shell with column density
$\sim 10^{20}~\cm^{-2}$ and expansion velocity $\sim 100~\kms$
we find that the averaged degree of polarization of the emergent
Ly$\alpha$ line photons reach as high as 0.05 when $\mu = 0.5$.

There are three interesting classes of primeval objects showing P-Cygni type 
profiles; the DLA candidate galaxies 
including DLA~2233+131 
\markcite{djor96, lu96, lu97} (Djorgovski et al. 1996, Lu et al. 1996, 1997) 
and DLA~2247-021 \markcite{djor97} (Djorgovski 1997), the Lyman break galaxies
at $3<z<4$ \markcite{ste96, ste97, low97} 
(Steidel et al. 1996, 1997, Lowenthal et al. 1997),
and the remote galaxies observed in the gravitational lens surveys
\markcite{fra97, fry97, tra97}
(Franx et al. 1997, Frye et al. 1997, Trager et al. 1997).

Firstly, several DLA galaxies with $3<z< 4$
are listed including DLA~2233+131 with $z=3.15$ by 
\markcite{djor96, djor97} Djorgovski et al. (1996,1997)
who concluded that these objects are progenitors of normal disk galaxies today.

In view of the point that the galactic rotation would erase 
the P-Cygni structure, the Ly$\alpha$ emitters or the giant H~II regions 
might be concentrated in a compact region.
It is noted that they also exhibit the P-Cygni type Ly$\alpha$ 
profiles, which is one of prominent characteristics of
nearby dwarf starbursts
\markcite{meu95, kun96, leg97, lei97} 
(e.g. Meurer et al. 1995, Kunth et al. 1996, Legrand et al. 1997, 
Leitherer 1997).
According to \markcite{meu95} Meurer et al. (1995),
there is evidence for existence of a dust lane 
or a thick H I disk component partly
obscuring the UV sources or the super star clusters.

Secondly, interesting astronomical objects showing P-Cygni type 
profiles are the Lyman break galaxies at $z \sim 3$ 
discovered by \markcite{ste96} Steidel et al. (1996) 
\markcite{low97} (see also Lowenthal et al. 1997).

From $\sim 75\%$ of the Lyman break galaxies obtained from their recent 
survey, \markcite{ste97} Steidel et al. (1997) detected Ly$\alpha$ emission 
lines which are weaker than expected from the UV continuum luminosities.
They ascribed this to the resonant scattering \markcite{kun98} 
(see Kunth et al. 1998).

The third group includes the remote galaxies which 
can be observed using the gravitational lens effect of 
the clusters of galaxies \markcite{fra97, fry97, tra97} 
(Franx et al. 1997, Frye et al. 1997, Trager et al. 1997).
Majority of the obtained profiles show the prominent P-Cygni type.

One other important application may be found in the first generation 
star clusters investigated by \markcite{hai97a, hai97b} Haiman and Loeb (1997a,b).
These primeval objects are expected to be observed at the redshift
of $5<z<10$ and therefore the Ly$\alpha$ emission will be located
most possibly in the near IR band where
the NGST can be used. The neutral hydrogen
layers enveloping the super star clusters may trace the Hubble-type expansion,
and the mechanism of the Ly$\alpha$ line formation may be studied in a
similar way introduced in this work \markcite{AL98} (Ahn \& Lee 1998).

We conclude that the Ly$\alpha$ lines emerging
from an expanding and optically thick medium are measurably 
polarized by up to 5 per cent and therefore a good constraint
on theoretical models is provided by band polarimetry.

\acknowledgments
 
We thank Dr. Leitherer and Dr. Meurer for kindly providing their
preprints and reprints.
We are very grateful to Prof. Blandford for
his invaluable comments about this work.


\clearpage
 
\begin{figure}
\caption{Linear degree of polarization (points with error bars), 
emergent flux (thick solid lines), and polarized flux (thin solid lines)
from linearly expanding slabs 
with $\tau_0=10^2, \ 10^3, \ 10^4$ and $s_{max}=1, \ 10$.  \label{fig1}}
\end{figure}

\begin{figure}
\caption{
Linear degree of polarization (points with error bars),
scattered flux (thick solid lines), and polarized flux (thin solid lines)
from a hemispherical shell that is made of local slabs 
with $\tau_0=10^4$ and $s_{max}=10$. The obscuration 
by an opaque component extends from the equatorial part 
to the region of colatitude
$\theta= 0^\circ,\ 30^\circ, \  60^\circ$. 
The viewing angle is chosen so that the averaged degree 
of polarization is maximal.  \label{fig2}}
\end{figure}

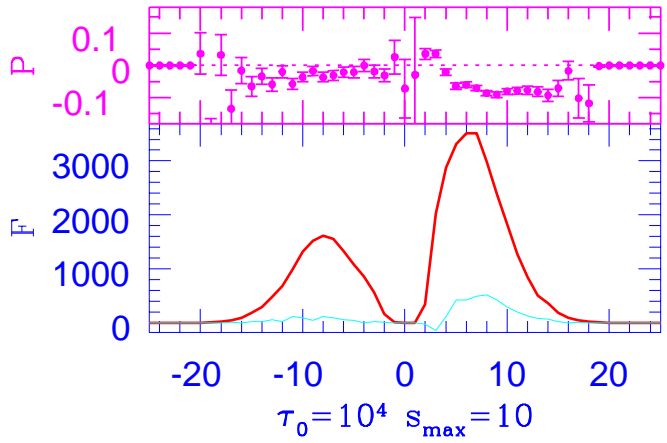
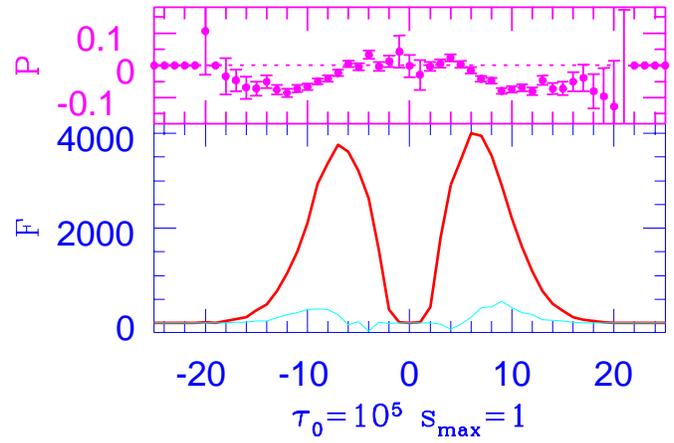
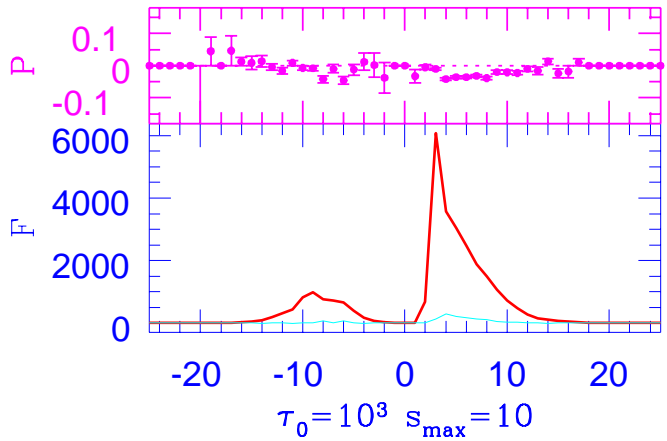
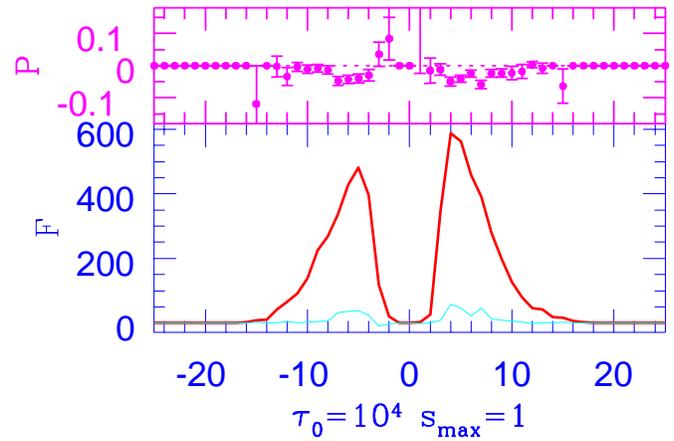
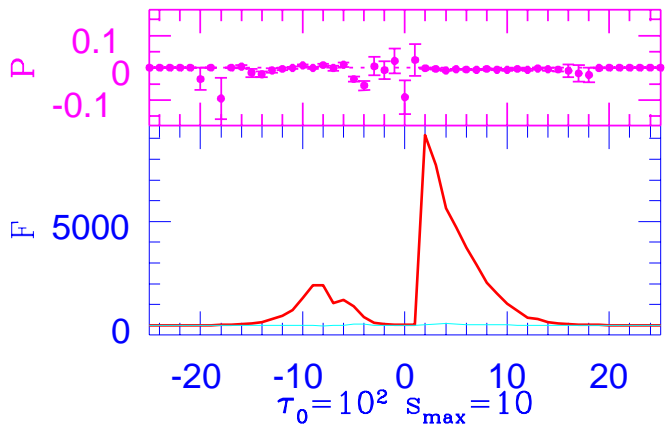
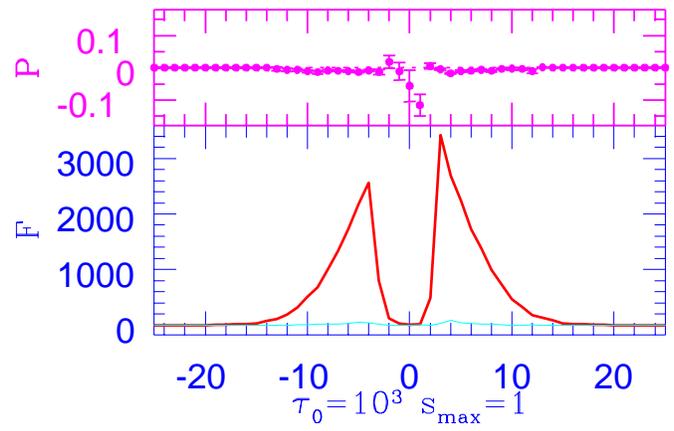
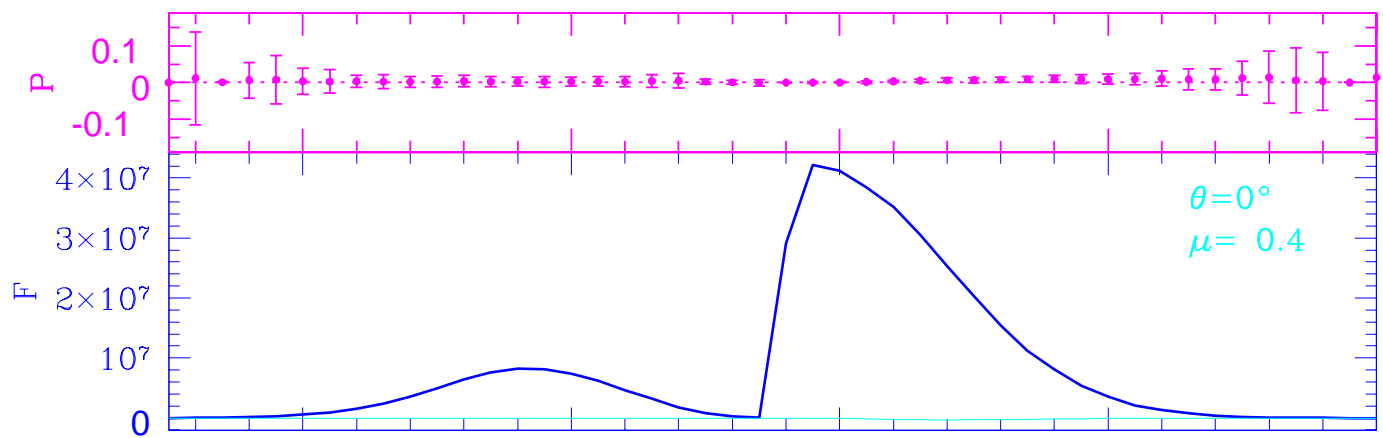
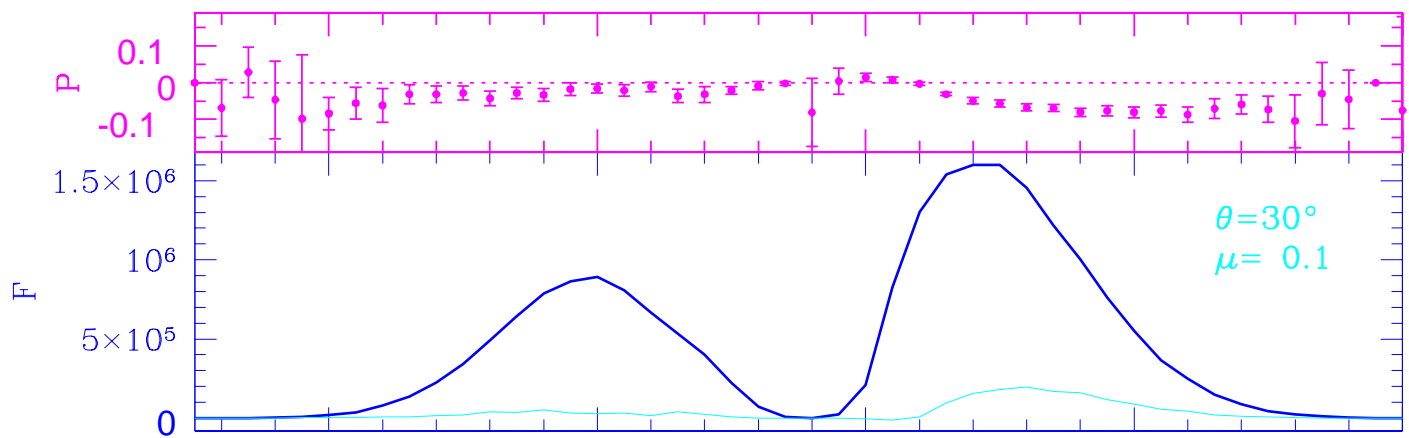
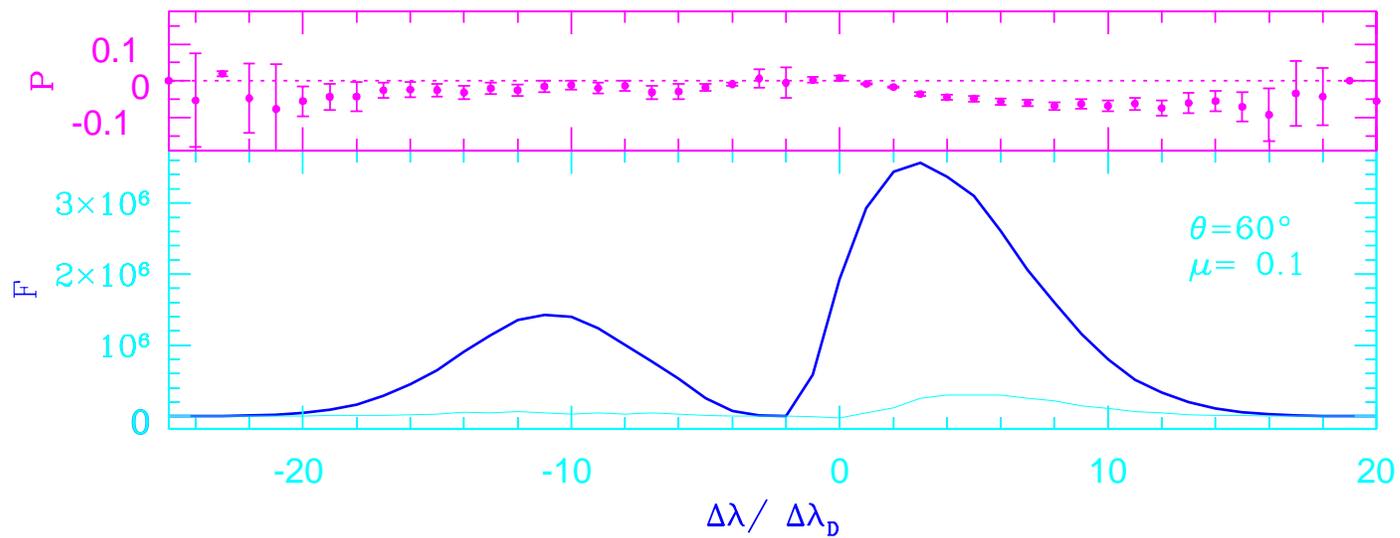
\end{document}